\documentclass[intlimits,twoside,a4paper]{article}
\usepackage[cp1251]{inputenc}

\usepackage{cmpj3}

\usepackage{bm}

\newcommand*{\VL}{V_\mathrm{L}}
\newcommand*{\Er}{E_\mathrm{r}}
\newcommand*{\EF}{E_\mathrm{F}}

\newcommand*{\Nc}{N_\mathrm{c}}
\DeclareMathOperator{\diag}{diag}
\usepackage{xcolor}

\definecolor{emerald}{RGB}{47,161,146}
\newcommand\hl[1]{%
	\bgroup
	\hskip0pt\color{emerald!90!black}%
	#1%
	\egroup
}

\issue{2018}{21}{3}{33703}
\doinumber{10.5488/CMP.21.33703}

\title{Impurity induced broadening of Drude peak in strained graphene}
\author{V.O. Shubnyi\refaddr{label1}, S.G. Sharapov\refaddr{label2}, Y.V. Skrypnyk\refaddr{label2}}
\addresses{
	\addr{label1} Department of Physics, Taras Shevchenko National University of
	Kyiv, \\6 Academician Glushkov Ave., 03680 Kyiv, Ukraine
	\addr{label2} Bogolyubov Institute for Theoretical Physics of the National Academy of
	Science of 	Ukraine, \\14-b Metrologichna St., 03680 Kyiv, Ukraine
}

\date{Received May 24, 2018}

\begin{document}
	\maketitle
	
\begin{abstract}
		The recent experimental study of the far-infrared transmission spectroscopy of
		monolayer graphene has shown that the Drude peak width increases by more than
		$10\%$ per $1\%$ of applied mechanical strain, while the Drude weight remains unchanged. We study the influence of the strain on the
		resonant impurity scattering. We propose a mechanism of augmentation of
		the scattering rate due to the shift in the position of resonance. Using the
		Lifshitz model of substitutional impurities, we investigate changes in the Drude peak weight and width as functions of the Fermi energy, impurity concentration and magnitude of the impurity potential.

\pacs 72.80.Vp, 62.20.-x, 72.80.Vp, 81.05.ue, 73.22.Pr, 71.55.Ak	

\end{abstract}
	
	\section{Introduction}
	This work is dedicated to 80th birthday of our colleague and a remarkable
	physicist Professor Ihor Stasyuk. It is a pleasure for us to present a recent result on
	strained graphene. We study intricate physics related to the interplay between
	the band structure and strain using the methods of the Green's function approach,
	to which I. Stasyuk has made a significant contribution \cite{Stasyuk}.
	
	Outstanding mechanical strength and stiffness of graphene are marked by its
	capability to sustain reversible elastic deformations in excess of $20\%$
	\cite{Lee2008}. These features are related to in-plane $\sigma$ bonds
	formed by hybridized $2s$, $2p_x$ and $2p_y$ electron orbitals of carbon atoms in
	the lattice. The remaining $2p_z$ orbitals form valence and conductance bands responsible for the observed transport and
	optical properties of graphene (for a review, see, e.g. \cite{CastroNeto2009}).
	
	In the simplest case, the strain is uniaxial. This strain configuration has been investigated by quite a few authors. In particular, a number of theoretical studies \cite{Pellegrino2010prb, Pereira2010epl, Pellegrino2011prb,
		Pellegrino2011hpr, Oliva-Leyva2014jpcm}
	predicted that the optical conductivity
	changes anisotropically with respect to the direction of the strain. These
	predictions seem to be in agreement with the transparency measurements in the
	visible range made on large-area chemical vapor deposited (CVD) monolayer
	graphene pre-strained on a polyethylene terephthalate (PET) \cite{Ni2014}.
	
	The recent study in optical conductivity response to the strain
	\cite{Chhikara2017} revealed some unexpected features in the far-infrared
	transmission spectra of a CVD monolayer graphene on PET substrate. It was found
	that the Drude peak width increases by more than $10\%$ per $1\%$ of the applied
	uniaxial strain, while the Drude weight remains unchanged.
	
	The purpose of the present work is to study the impact of strain on the optical conductivity in the presence of point defects. These defects are assumed to be either chemically substituted
	carbon atoms including their absence, i.e., vacancies, or adsorbed atoms or
	molecules on a graphene sheet. They may originate as a byproduct of a
	fabrication method, or can be purposefully superimposed on the graphene sheet by
	exposure to an active environment. In either case, one should expect a finite
	amount of defects on  graphene \cite{Ni2010}.
	
	In \cite{Chhikara2017}, contribution of the point defects to the Drude peak width was considered within the Born approximation. It was
	found that the resulting variation of the Drude width under the
	strain is insignificant. It is known, however, that certain point defects may
	lead to the appearance of the resonance impurity states. The resonance energy can be
	located near the Dirac point, in which case the resonance is well defined. This makes properties of the system, such as
	optical conductivity, very sensitive to the position of the Fermi
	energy $\EF$ relative to the resonance energy. Even a small variation of this parameter
	can lead to a dramatic increase in the Drude width.
	
	The paper is organized as follows. We begin by presenting in section~\ref{sec2} the Hamiltonian of the electronic subsystem in strained graphene. To describe impurities, we employ the Lifshitz model. In section~\ref{sec3}, we use the Green's function formalism to calculate the electron self-energy function within the average T-matrix approximation. Then, we use the self-energy function to calculate the Drude weight and width. In section~\ref{sec4}, we present the results for various impurity concentrations and impurity perturbations, and analyze conditions for which the used approximations are valid.
		
	Throughout the paper, units $\hbar=e=1$ are chosen. Particularly, we will omit $e^2/\hbar^2$ factor in conductivity and the Drude weight (the last one is thus measured in units of energy).
	\section{Model} \label{sec2}
	It is assumed that electrons in
	impure graphene can be described by the Hamiltonian of separable form
	\begin{equation}
	{\hat H} =  {\hat H}_{0} + {\hat V},
	\label{eqn:h}
	\end{equation}
	with the first term characterizing the electronic subsystem of a clean graphene
	sheet, and the second term adding impurity perturbation to the model.
	\begin{figure}[!b]
		\centering
		\includegraphics[width=0.4\textwidth]{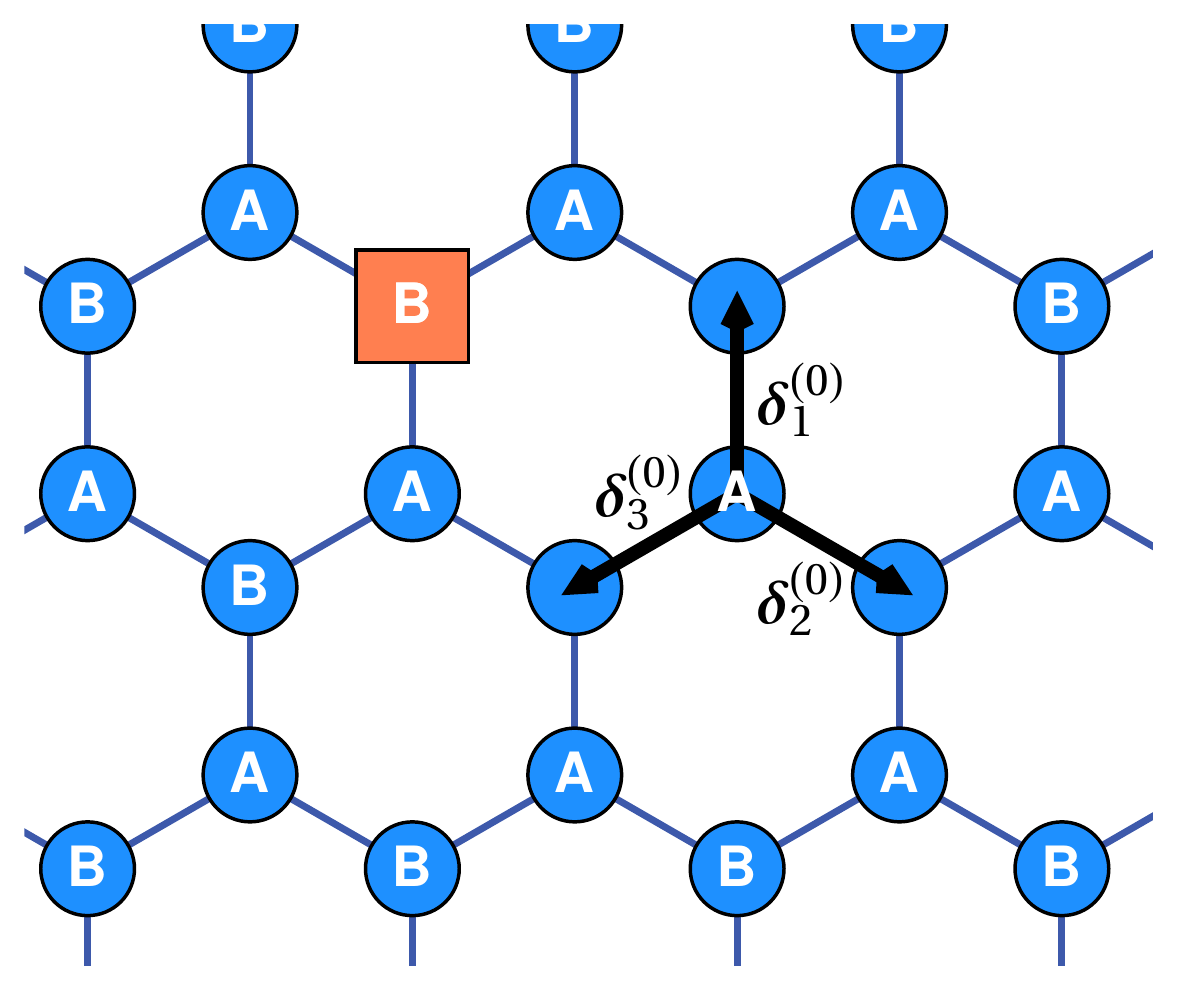}
		\caption{(Colour online) Schematic representation of the undeformed impure graphene crystal. Blue circles represent the host atoms, and an orange square stands for a substitutional impurity. The letters A and B label sites on two sublattices. The nearest neighbour vectors $\bm \delta_n^{(0)}$ are
			$\bm \delta_1^{(0)} =  a(0,1)$,
			$\bm \delta_2^{(0)} =  a(\sqrt{3}/2,-1/2)$
			and
			$\bm \delta_3^{(0)} =  a(-\sqrt{3}/2,-1/2)$.}
		\label{fig:lattice}
	\end{figure}
	
	The crystal lattice for an undeformed sample with impurities is depicted
	in figure~\ref{fig:lattice}. We assume that the substitutional
	impurities do not
	change the lattice structure. For a clean graphene crystal, the
	hexagonal
	structure has two non-equivalent sublattice positions A and B for each
	Bravais lattice cell. We use the second quantization formalism, with the creation (annihilation) operators $\hat c_{\alpha i}^+$
	($\hat c_{\alpha i}^{}$) which add (remove) the electronic state at the $i$-th
	cell on
	the $\alpha = \text{A}, \text{B}$ sublattice. Usually, one works in the nearest-neighbour
	approximation, where the only non-zero hopping integrals are those between
	the nearest neighbours A and B. Therefore, the resulting form of the
	Hamiltonian for the hexagonal lattice is:
	\begin{equation}
 {\hat H}_0 = - \sum_{\langle ij \rangle}\big( t_{ij}^{}
	\hat c_{\text{A} i}^+ \hat c_{\text{B} j}^{\phantom{+}} + t_{ij}^\ast
	\hat c_{\text{B} j}^+ \hat c_{\text{A} i}^{\phantom{+}}\big),
	\label{eqn:h0}
	\end{equation}
	where the sum $\sum_{\langle ij \rangle}$ goes over the closest Bravais cells
	$i$ and $j$.
	Here, $t_{ij}$ are the hopping integrals for the nearest-neighbouring atoms
	with indices $(i,\text{A})$ and $(j, \text{B})$.
	In the case of unstrained graphene, the hopping integrals are position and direction independent,
	$t_{ij} = t$.
	The value of the hopping integral is usually estimated as $t \approx 2.7$~eV \cite{Andrei2012RPP}.
	As we turn on the strain, the distances between the atoms vary,
	and in general case, $t_{ij}$ no longer takes such a simple form.
	
	Since we consider a simple uniaxial strain, which is uniform along the direction of the applied stress, the components of two-dimensional strain tensor $\bar{ \pmb{\varepsilon}}$
	are independent of equilibrium positions of atoms~$\mathbf{r}$. Accordingly, the displacement vector $\mathbf{u}(\mathbf{r})$ reads
	$\mathbf{u}(\mathbf{r}) = \bar{\pmb{\varepsilon}} \cdot \mathbf{r}$. Thus, the actual position of an atom
	$\mathbf{r}^\prime =\mathbf{r} +\mathbf{u}(\mathbf{r})$ can be written as
	$\mathbf{r}^\prime = (\bar{ \mathbf{I}} + \bar{\pmb{\varepsilon}}) \cdot \mathbf{r}$, where $\bar{ \mathbf{I}}$ is the unit
	$2 \times 2$ matrix.

	The strain tensor acquires the diagonal form
	$\bar{\pmb{\varepsilon}} = \diag( \varepsilon, - \nu \varepsilon )$
	in the coordinate system aligned with the axis of the applied stress.
	Here, $\varepsilon$ is the
	relative longitudinal expansion, and $\nu$ is the Poisson's ratio
	($\nu=0.16$ in graphite and may possibly be
	even smaller in graphene on substrate \cite{Chhikara2017}).
	Note that the planar deformation of the hexagonal crystal in the basal plane is determined by the two
	independent stiffness (compliance) tensor components, viz. it behaves as an isotropic planar solid \cite{Landau7}.

	The three vectors that connect atom A with the nearest
	neighbouring atoms B, $\bm \delta_n$ ($n=1,2,3$), change from their equilibrium value $\bm \delta_n^{(0)}$ (see
	figure~\ref{fig:lattice}) independently of the cell index: $\bm \delta_n = (\bar{ \mathbf{I}} + \bar{\pmb{\varepsilon}}) \cdot
	\bm \delta_n^{(0)}$. Accordingly, the
	hopping integral does not depend on the cell index $i$ of the A atom, but only on the direction to the neighbours.
	Thus, we end up with the three distinct hopping
	integrals, which we will denote as $t_{\{n\}}$.
	It is usually assumed that the change in the overlap integrals
	of $p_z$-orbitals \cite{Suzuura2002}
	can be adequately described by the first order expansion in strain (see also  in \cite{Naumis2017RPP}):
	\begin{equation}
	t_{\{n\}} \approx t - \frac{\beta t}{a^2} \Big( \bm \delta_n^{(0)} \bar{\pmb{\varepsilon}} \bm \delta_n^{(0)} \Big),
	\label{eqn:t-n}
	\end{equation}
	where $a \approx 1.42$~{\AA}  is the equilibrium distance between the
	carbon atoms, and $\beta = - \rd \ln t / \rd \ln |\delta| \sim 3 {-} 4$ is the dimensionless
	Gruneisen parameter.
	
	It can be shown that under uniform uniaxial strain, the density of states (DOS) per unit cell and spin
	(including the valley degeneracy)
	acquires the form
	\cite{Oliva-Leyva2014jpcm,DeJuan2013a,Shah2013}
	\begin{equation}
	\rho_\varepsilon (E) =\frac{2 |E|}{W^2_\varepsilon}\,, \qquad |E| \ll t,
	\label{eqn:dos-0}
	\end{equation}
	where $W_\varepsilon$ is the strain dependent bandwidth,
	\begin{equation}
	W^2_\varepsilon = [1 -  \beta (1-\nu) \varepsilon] W^2_0.
	\label{eqn:w-s}
	\end{equation}
	Here, $W_0 = (\sqrt{3} \piup )^{1/2} t  \simeq 2.33 t$ is
	the effective bandwidth or the  energy cutoff for unstrained graphene that preserves the number of states
	in the Brillouin zone. It is worth mentioning that energy is counted from the Dirac point.
	In what follows, we will assume that $\nu=0$, because
	if necessary it can be easily restored by replacing $\varepsilon \to \varepsilon (1 -\nu)$.
	
	In the simplest model of substitutional impurities, widely referred to as the
	Lifshitz model \cite{Lifshitz1964}, we add the constant potential $\VL$ to the sites occupied by the
	impurity atoms. This leads to the diagonal term in the Hamiltonian
	\begin{equation}
	{\hat V} = \VL \sum_{l \alpha} \eta_{\alpha l} \hat c_{\alpha l}^+
	\hat c_{\alpha l}^{\phantom{+}}\,,
	\end{equation}
	where $\eta_{\alpha l}$ equals one on the impurity sites, and zero otherwise. For a lattice with $N$ atoms with the impurity concentration $c$, the total number of impurity sites is $cN$, but the distribution of impurities is not specific, and differs from one sample to another. This model can also be used to describe vacancies in a graphene sheet deposited on a substrate. Additionally, it can describe adsorbed atoms, provided that the hopping integral between the additional orbital of the adsorbed atom or molecule, and the $p_z$ orbital of the carbon atom of the host is larger than the bandwidth $W_0$.
	\section{Formalism}\label{sec3}
	
	The problem of inclusion of impurities in the electronic structure can be
	treated by using the Green's function method (see \cite{Yonezawa1968,Elliott1974,Ehrenreich1976}
	for a general review, or \cite{Pershoguba2009} for a graphene-specific application). The Green's function is related
	to the Hamiltonian (\ref{eqn:h}) as ($\hat{\mathbf I}$ is the unit matrix):
	\begin{equation}
	\bm{\hat\mathcal{ G}} (E) = (E  \hat{\mathbf I} -  \hat{\mathbf H}_0 -  \hat{\mathbf V} )^{-1}.
	\label{eqn:greens-function}
	\end{equation}
	We prefer to work in the site representation, where both the Hamiltonian and the Green's function are $N \times N$ matrices in the basis of one-particle states $\hat c^+_{\alpha l} |0 \rangle$ (here $|0 \rangle$ is the vacuum state).
	
	In a general case, this Green's function matrix has a complicated structure due to irregularity in positions of the
	impurities. To treat this problem,
	we are averaging the Green's function over all possible configurations of
	impurities at a fixed concentration $c$:
	$\hat{\mathbf G} =\langle \bm{\hat\mathcal{ G}} \rangle$. It turns out that with an increasing number of atoms in the lattice, the Green's function $\bm{\hat\mathcal{ G}}$
	approaches the average value $\hat{\mathbf G}$ \cite{LifshitsGredeskulPastur1982}.
	The averaging reestablishes
	the translational invariance, and the averaged Green's function $\hat{\mathbf G}$ can be related to the
	host Green's function
	\begin{equation}
	\hat{\mathbf g} = (E \hat{\mathbf I} - \hat{\mathbf H}_0)^{-1}
	\label{eqn:g-small}
	\end{equation}
	by means of the Dyson equation:
	\begin{equation}
	\hat{\mathbf G} = \hat{\mathbf g} + \hat{\mathbf g} \hat {\bm \Sigma} \hat{\mathbf G}.
	\end{equation}
	If we take into consideration only the single-site scattering, the self-energy operator $\hat {\bm\Sigma}$ becomes diagonal. Omission of the cluster scattering leaves us with the
	scalar function $\varSigma = \varSigma(E)$, expressed via $\hat{\bm\Sigma} = \varSigma \hat{\mathbf I}$. Therefore, the Green's
	function $\hat{\mathbf G}$ can be expressed via $\hat{\mathbf g}$ as
	\begin{equation}
	\hat{\mathbf G}(E) \approx \hat{\mathbf g} [ E - \varSigma (E) ].
	\end{equation}
	
	In the present work we employ the average T-matrix
	approximation (ATA), in which $\varSigma(E)$ is expressed via the diagonal
	element of the host Green's function in the site representation $g_0(E)=
	g_{\alpha i, \alpha i}(E)$ as follows:
	\begin{equation}
	\varSigma_{(\mathrm{ATA})} (E) = \frac{c \VL}{1 - (1 - c) \VL
		g_0(E)}.
	\label{eqn:sigma-ata-lifshitz}
	\end{equation}
	To obtain an analytical expression for $\varSigma$ to work with, we derive
	$g_0$ via the relation between it and the density of states per unit cell
	$\rho_0(E)$:
	\begin{equation}
	\label{g-0}
	2 \Im g_0 (E) = - \piup \rho_0(E).
	\end{equation}
	This determines the imaginary part; the real part is restored using the
	Kramers-Kronig relation. The resulting expression for $g_0$ is:
	\begin{equation}\label{g0}
	g_{0} (E)  = \frac{E}{W^2_\varepsilon}
	\ln\left(\frac{E^2}{W^2_\varepsilon}\right) - \ri \frac{\piup
		|E|}{W^2_\varepsilon} \,, \qquad |E| \ll t.
	\end{equation}
	
	We now discuss how the the calculated self-energy $\varSigma_{(\mathrm{ATA})}$ is related to
	the spectroscopy measurements \cite{Chhikara2017}.
	One of the advantages of the infrared spectroscopy
	as compared to the DC transport measurements is that
	it allows one to find independently both the Drude spectral weight and the optical scattering rate.
	We assume that the real part of the Drude conductivity is of a Lorentzian shape
	\begin{equation}
	\label{conductivity}
	\Re \sigma(\omega) = \frac{D (E_{\text F})}{\piup} \frac{\Gamma (E_{\text F})}{\omega^2 + \Gamma^2 (E_{\text F})}\,,
	\end{equation}
	where $D$ is the Drude weight and $\Gamma$ is the Drude peak width or optical scattering rate.
	Both these quantities depend on the Fermi energy.
	The Drude weight can be related to the dc limit, $\sigma_{\mathrm{dc}} =
	\Re \sigma(\omega =0) = D/ (\piup \Gamma )$, of the dynamical conductivity (\ref{conductivity}).
	
	According to the Matthiessen's
	rule, the total optical scattering rate is
	$\Gamma = 2 \sum_i \Gamma_i$, where $\Gamma_i$ are the contributions from the different channels of
	single particle scattering, e.g., short-range point defects, long-range charged impurities, acoustic phonons, surface phonons in
	the substrate and grain boundaries. Each of these contributions may be strain dependent.
	As was mentioned above, in the present work we restrict ourselves to the scattering by point impurities:
	\begin{equation}
	\Gamma (E_{\text F}) = - 2 \Im \varSigma (E_{\text F}).
	\label{eqn:gamma-im}
	\end{equation}
	
	Although the uniaxial strain makes the Drude weight anisotropic \cite{Chhikara2017},
	we will only use the Drude weight averaged over strain directions. Thus,
	using the expression for the dc conductivity for graphene
	obtained in the bare bubble approximation at zero temperature \cite{Kumazaki2006},
	we obtain
	\begin{equation}
	D(E_{\text F}) = \frac{2}{\piup }\left [1 + \frac{E_{\text F} -
		\Re\varSigma(E_{\text F})}{-\Im\varSigma(E_{\text F})} +
	\frac{-\Im\varSigma(E_{\text F})}{E_{\text F} -
		\Re\varSigma(E_{\text F})} \right]
	\arctan \left[\frac{E_{\text F} -
		\Re\varSigma(E_{\text F})}{-\Im\varSigma(E_{\text F})} \right] \left[-\Im\varSigma(E_{\text F}) \right].
	\label{eqn:d}
	\end{equation}
	For small concentrations of impurities, when inequality $|\varSigma(E)| \ll E$ holds, equation~(\ref{eqn:d})
	takes a rather simple form
	\begin{equation}
	D \approx E_{\text F} - \Re\varSigma(E_{\text F}).
	\label{eqn:d-simple}
	\end{equation}

	\section{Results and analysis}\label{sec4}
	
	\subsection{Strain dependence of the impurity resonance energy}
	
	Since a small impurity concentration, $c \ll 1$, is considered, it is safe 
	to neglect
	$c$ in the denominator of the ATA expression (\ref{eqn:sigma-ata-lifshitz}).
	Substituting the diagonal element of the
	host Green's function (\ref{g-0}) in equation~(\ref{eqn:sigma-ata-lifshitz}), we 
	obtain, for example, that the imaginary part of the self-energy reads
	\begin{equation}
	\Im \varSigma(E) \approx \dfrac{-c \VL^2
		\piup |E|/W^2_\varepsilon}{\left[1-\dfrac{\VL E}{W^2_\varepsilon}
		\ln\left(\dfrac{E^2}{W^2_\varepsilon}\right)  \right]^2 + \left[\dfrac{\piup \VL
			E}{W_\varepsilon^2}\right]^2}.
	\label{eqn:gamma-lifshitz-simple}
	\end{equation}
	The denominator of (\ref{eqn:gamma-lifshitz-simple}) has two terms, both of which are non-negative.
	While the last term is a smooth function of energy, the first term can be zero. Therefore, from the last equation it is clear that the impurity scattering rate (\ref{eqn:gamma-im}) should have a maximum at some energy.
	For $|\VL| \gg W_\varepsilon$
	 the location of this maximum~$\Er$ can be approximated by the solution of the well-known Lifshitz equation, $1 = \VL \Re g_0(\Er)$, or
	\begin{equation}
	\label{eqn:lifshitz-equation}
	1 -\frac{\VL \Er}{ W^2_\varepsilon} \ln\left( \frac{\Er^2 }{ W^2_\varepsilon} \right)  = 0.
	\end{equation}
	For $\VL  < 0$, the position of the extremum is above the Dirac point, $\Er >0$, and vice versa. This unusual positioning property
	holds true for a general spectrum consisting of two symmetric bands touching each other at a single point \cite{Skrypnyk2007FNT}. Without compromising generality of our consideration, we
	will consider the case $\VL < 0$, and the electron-doped graphene $\EF > 0$.
	
	To find the solution of equation~(\ref{eqn:lifshitz-equation}),
	one can use a recursive procedure:
	\begin{equation}
	\Er^{(n)} = \frac{W_\varepsilon^2}{2 (-\VL)} \frac{1}{\ln (W_\varepsilon/|\Er^{(n-1)}|)}\,, \qquad \Er^{(1)}
	= \frac{W_\varepsilon^2}{2 (-\VL)}\,,
	\label{eqn:lifshitz-eqn-solution-iteration}
	\end{equation}
	with $\Er^{(n)}$, $n \to \infty$ converging to the exact solution $\Er$. Alternatively, one can express $\Er$ in terms of 
	the inverse function $\mathcal{W}(z)$ defined as a solution to the equation $z=\mathcal{W} \exp(\mathcal{W})$, 
	also known as the Lambert $\mathcal{W}$-function. This function has two branches $\mathcal{W}_0(z)$ and $\mathcal{W}_{-1}(z)$, 
	which represent separate solutions of the equation. Specifically, we will use the $\mathcal{W}_{-1}$ branch, 
	which gives $|\mathcal W_{-1}(x)| > 1$ for $x < 0$. In terms of this function,
	\begin{equation}
	\Er = \frac{W_\varepsilon^2}{2 \VL} \frac{1}{\mathcal{W}_{-1} (W_\varepsilon/2\VL)}.
	\label{eqn:lifshitz-eqn-solution-exact}
	\end{equation}
	
	This extremum at $\Er$ can be regarded as a well-defined impurity resonance when its width is much smaller than the difference between $\Er$ and the Dirac point energy. The corresponding condition $|\Im \varSigma(\Er)| \ll |\Er|$ gives us the following
	\begin{equation}
	\zeta = \frac{\piup}{2} \frac{1}{\ln (W_\varepsilon/|\Er|) - 1} \ll 1, \qquad |\Er| \ll W_\varepsilon.
	\label{eqn:zeta}
	\end{equation}
	According to (\ref{eqn:lifshitz-equation}), (\ref{eqn:lifshitz-eqn-solution-iteration}), (\ref{eqn:lifshitz-eqn-solution-exact}), small $|\Er|$ corresponds to large $|\VL|$. Thus, a well-defined resonance in the Lifshitz model is present only for $|\VL| \gg W_\varepsilon$. It should be understood that the Lifshitz model is, in a certain sense, phenomenological. More complicated impurity models that are more appropriate to the real physical situation solve the apparent problem of the excessively large impurity potential. At the same time, these models yield qualitatively similar results. We would like to stress that the Lifshitz impurity model is capable of providing a semi-quantitative description of experimental ARPES data on the impure graphene \cite{26}.
	
\begin{figure}[!t]
		\centering
		\includegraphics[height=2.4in]{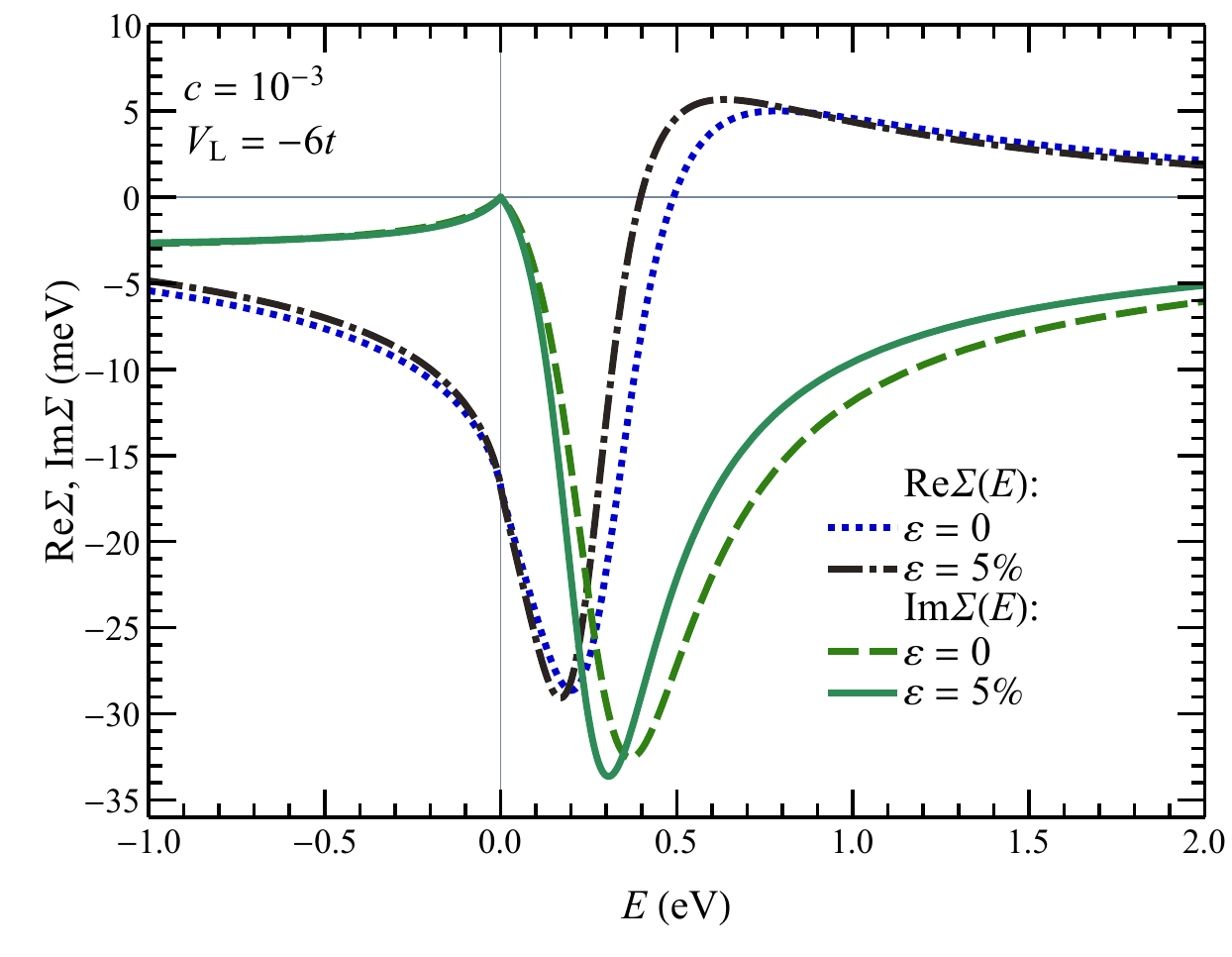}
		\caption{(Colour online) The self-energy $\varSigma(E)$ in the average T-matrix approximation in the Lifshitz model of impurities. $\Re \varSigma (E)$ and $\Im \varSigma(E)$ as the functions of energy $E$ for
			two values of strain $\varepsilon = 0$ and $\varepsilon = 5 \%$, calculated for $\VL = - 6 t$, $c =
			10^{-3}$.}
		\label{fig:sigma}
	\end{figure}

	We show in figure~\ref{fig:sigma} both $\Im \varSigma (E)$
	and $\Re \varSigma (E)$ computed using equation~(\ref{eqn:sigma-ata-lifshitz}) in the absence of strain and for $\varepsilon = 5\%$. 
	A moderate value $\VL = -6 t$ is chosen and the concentration of impurities is $c = 10^{-3}$. 
	Both the resonance dip in $\Im \varSigma(E)$ 
	and the corresponding S-shaped feature in $\Re \varSigma(E)$ are evident in figure~\ref{fig:sigma}. 
	These features are expressed in a narrow energy window. The span of the window corresponds to the above obtained width of the impurity resonance $\zeta \Er$.
	One can also see in figure~\ref{fig:sigma} that a small strain~$\varepsilon$ slightly shifts the
	position of the resonance towards the Dirac point. This shift, as we will show later, results in
	a considerable increase in the scattering rate $\Gamma$.
	
	Let us assume that the position of the resonance $\Er$ changes under the strain linearly:
	\begin{equation}
	\Er (\varepsilon) = (1 - \alpha \varepsilon) \Er (0),
	\label{eqn:er-s-linear}
	\end{equation}
	where $\Er(0)$ is the resonance energy in the absence of strain.
	Substituting equation~(\ref{eqn:er-s-linear}) to the Lifshitz equation (\ref{eqn:lifshitz-equation}), one can analytically estimate 
	the value of the coefficient $\alpha$:
	\begin{equation}
	\alpha = \frac{\lambda -1}{\lambda - 2}\beta , \qquad \lambda = -
	\frac{W_0^2}{\VL \Er (0)}.
	\label{eqn:er-fit}
	\end{equation}
	
\begin{figure}[!t]	
		\centering
			\includegraphics[width=0.47\textwidth]{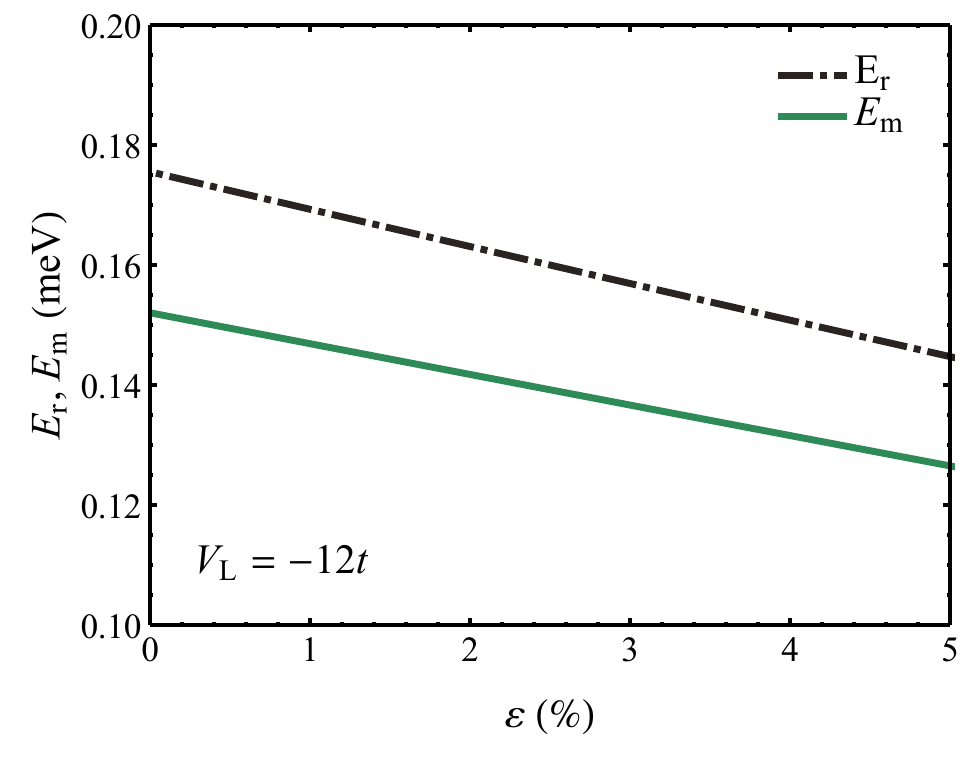}
			\includegraphics[width=0.5\textwidth]{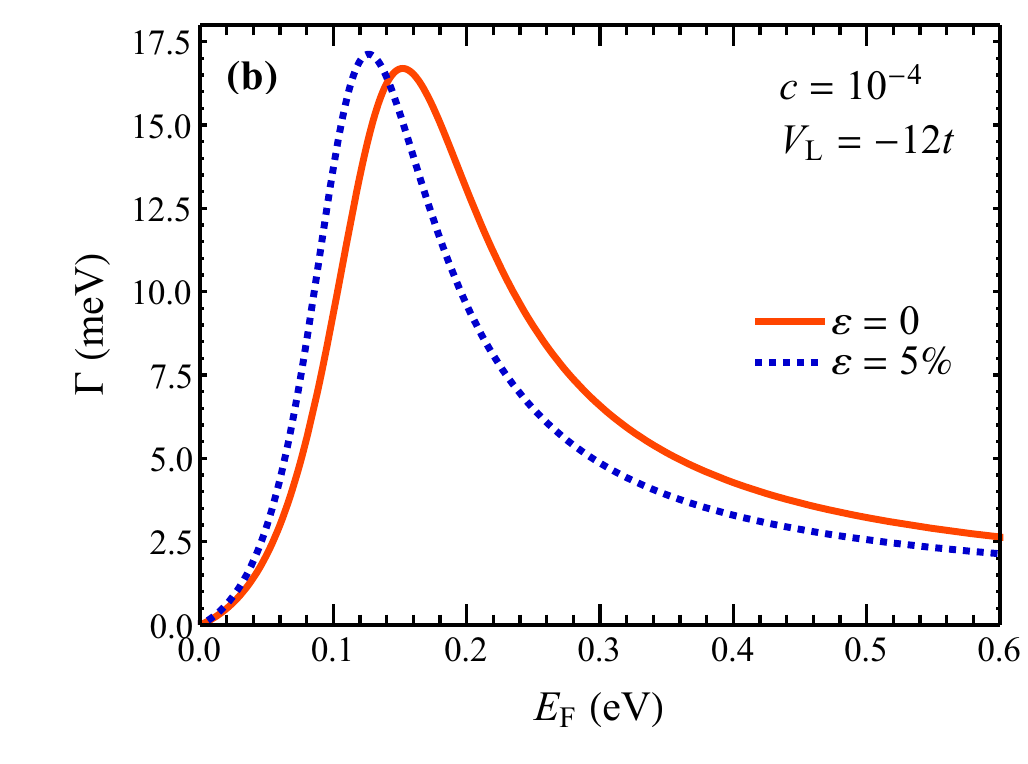}
		\caption{(Colour online) Resonance in the scattering rate $\Gamma$ calculated for $\VL = - 12t$, $c = 10^{-4}$. (a) The dependence of the resonance energy $\Er(\varepsilon)$ on the strain. 
			The solution of equation~(\ref{eqn:lifshitz-equation}) is shown by the dash-dotted (black) line.
			The dependence of the position
			of the actual minimum of the function $\Im \varSigma (E)$ on the strain $\varepsilon$ is shown by the
			solid (green) line. 
			(b) Drude peak width $\Gamma$ as a function of the Fermi energy $\EF$, 
			calculated for strain $\varepsilon = 0$ and $\varepsilon = 5 \%$.} \label{figure:2}
	\end{figure}

	In figure~\ref{figure:2}~(a), we show both the value of $\Er$ (dash-dotted black line) and the energy of the minimum~$E_\mathrm{m}$ of the function $\Im \varSigma (E)$ (solid green line)
	as functions of strain $\varepsilon$. The value $\VL = -12 t$ is used.
	One finds that the dependence $E_\mathrm{m}(\varepsilon)$ is significantly off
	the line $\Er (\varepsilon)$ discussed above. 
	However, the function $E_\mathrm{m}(\varepsilon)$ has a linear dependence similar to (\ref{eqn:er-s-linear}), but with a slightly different slope $\alpha'$. To obtain $\alpha'$ for the corresponding slope, one should substitute $\lambda$ in equation~(\ref{eqn:er-fit}) with a different parameter $\lambda' =-W_0^2/[\VL E_\mathrm{m}(0)]$, where $E_\mathrm{m}(0)$ is the value
	of the energy that gives the minimum at $\varepsilon = 0$. This difference between $\Er$ and $E_\mathrm{m}$ is connected to the fact that the condition for $\zeta$ (\ref{eqn:zeta}) is not decisively satisfied.
		
	Figure~\ref{figure:2}~(b) shows the resonance in $\Gamma$ as a function of 
	the Fermi energy $\EF$.
	The peak position is independent of the impurity concentration $c$. This allows one to tune the strength of the potential~$\VL$ to match the actual position of the resonance energy $\Er$ in the experimental sample.
	
	The imaginary part $\Im \varSigma(E)$ has an additional maximum at $E = 0$, 
	which corresponds to the Dirac point in the clean sample. In this point, equation 
	(\ref{eqn:gamma-lifshitz-simple}) yields $\Im \varSigma(0) = 0$ independent of strain. As it turns out, in the interval of the order $|c\VL|$ around the Dirac point, our approximation does not hold. An accurate description of the conduction band boundaries should include the renormalization of the propagator.
	
	\subsection{Small impurity concentrations}
	First, we analyze the case of those concentrations $c$ for which $|\varSigma(\EF)| \ll \EF$.
	This condition is satisfied for $c$ much smaller than some
	characteristic value $c_\chi$. The actual value of $c_\chi$ depends on $\VL$;
	the larger $|\VL|$ is, the smaller concentrations are required
	for deviations from the following analysis to show up.
	This concentration can be roughly estimated as $c_{\mathrm{\chi}} \sim \Er^2/W_0^2$ (see the sub-section~\ref{subsec:4.3} for a detailed discussion).
	
	\begin{figure}[!t]
		\centering
			\includegraphics[width=0.49\textwidth]{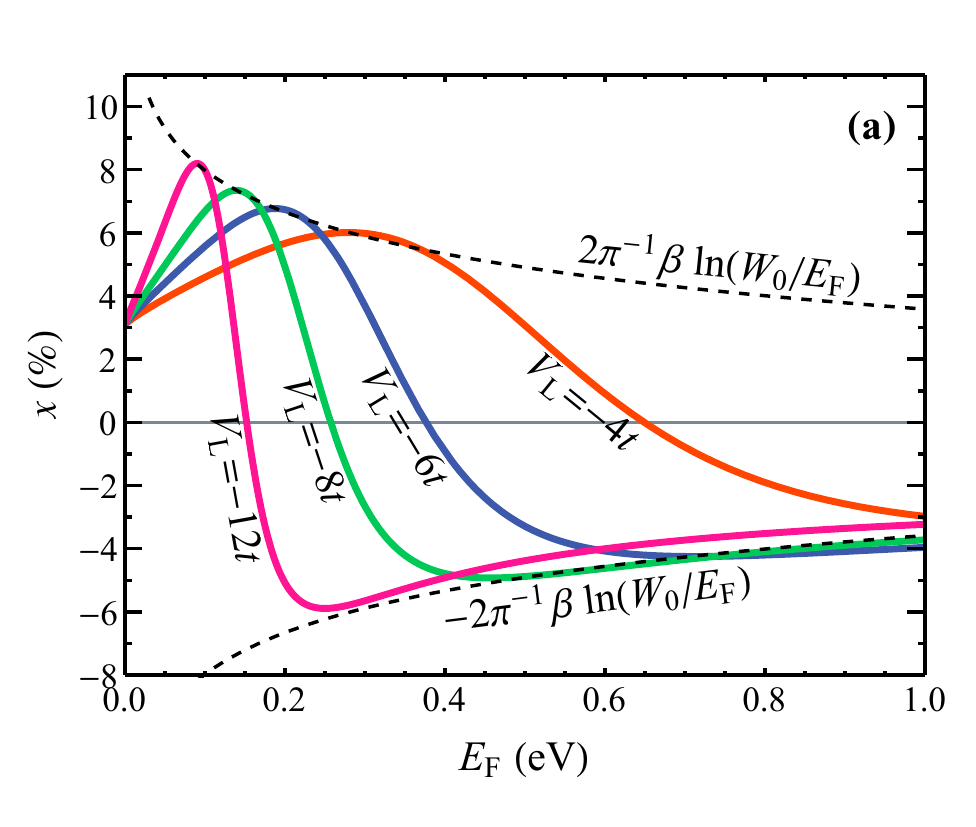}
			\includegraphics[width=0.49\textwidth]{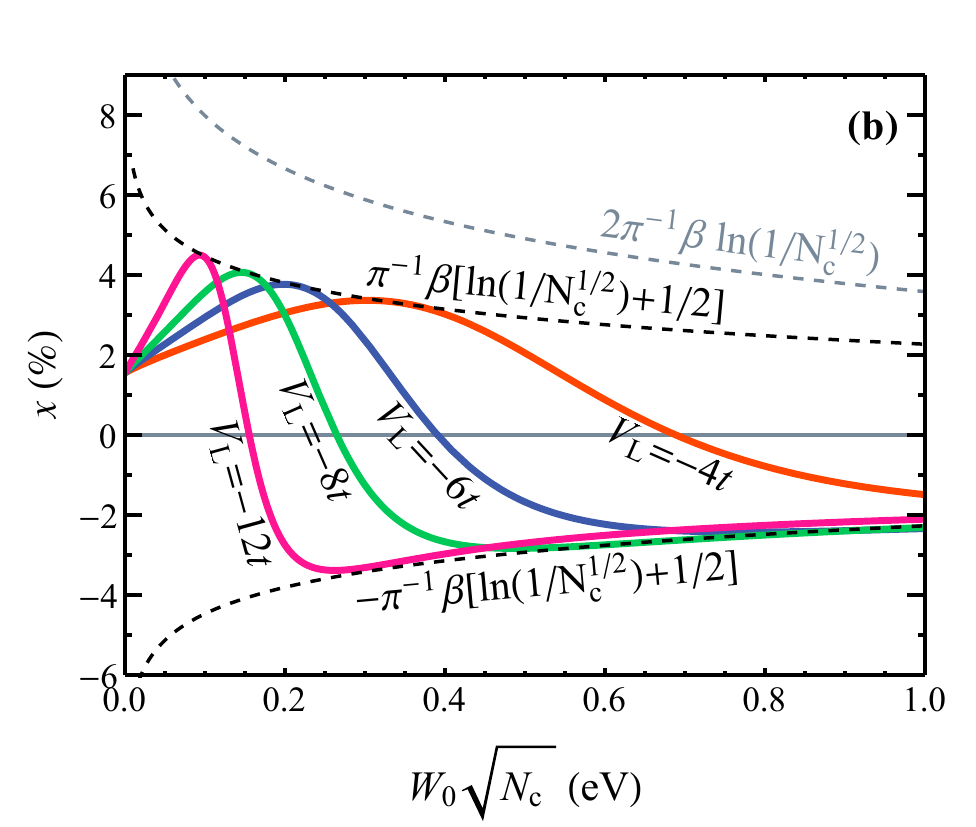}
		\caption{(Colour online) Relative gain $\varkappa$ in the Drude peak width $\Gamma$ under a $1 \%$ strain
			(a) as a function of the Fermi energy in the constant $\EF$ scenario;
			(b) as a function of $W_0 \sqrt{N_{\text c}}$ in the constant $\Nc = (\EF^\varepsilon/W_\varepsilon)^2$ scenario. }
		\label{fig:4}
	\end{figure}
	
With the strain $\varepsilon$, the Drude peak width
	$\Gamma(\EF)$ (\ref{eqn:gamma-im}) increases for $\EF$ residing below
	the resonance energy. The relative increase of the Drude width can be quite substantial, especially for values of the
	Fermi energy that lies at the half-width of the peak. Figure~\ref{figure:2}~(b) evidently demonstrates this feature. The gain turns out to be linear in strain, and can be described by the increment $\varkappa$, which gives a relative increase of the scattering rate
	$\Gamma$ per one percent of the strain $\varepsilon$:
	\begin{equation}
	\varkappa
	= \frac{\Gamma(\varepsilon) - \Gamma(0)}{\varepsilon \Gamma(0)}
	\approx \frac{1}{\Gamma(0)} \frac{\partial \Gamma(\varepsilon)}{\partial \varepsilon}.
	\label{eqn:kappa-epsilon}
	\end{equation}
	Figure~\ref{fig:4}~(a) shows $\varkappa$
	as a function of the Fermi energy $\EF$ calculated for various values of $\VL$.
	For a given value of the impurity perturbation $\VL$, the value of $\varkappa$ is independent of the impurity concentration $c$ for $c \ll c_\chi$.
	
	We can easily estimate the maximum gain
	$\varkappa_{\mathrm m}$ for a given impurity perturbation $\VL$. First, let us assume that $\EF$ remains constant
	throughout the process of deformation. The results for this scenario are
	shown in figure \ref{fig:4}~(a). As can be seen in figure~\ref{figure:2}~(b), the maximum values of $\varkappa(\EF)$ are reached when
	the Fermi energy lies at the half-width of the resonance. This happens
	when the first
	and the second term in the denominator of (\ref{eqn:gamma-lifshitz-simple}) are
	equal. Assuming that $\VL$ can be chosen arbitrarily, the maximum gain for a
	given $\EF$ can be estimated analytically as
	\begin{equation}
	\varkappa_{\mathrm m}^{(1)}
	\approx \dfrac{2}{\piup} \beta\ln\left( \dfrac{W_0}{|\EF|}\right).
	\label{eqn:kappa-m1}
	\end{equation}
	This can be seen in figure~\ref{fig:4}~(a), where the
	curves for various $\VL < 0$ are enveloped by the dashed curve described by
	(\ref{eqn:kappa-m1}). It turns out that the relative increase in $\Gamma$ is
	independent of impurity concentration.
	
	Apart from observing the substantial increase of the Drude width for
	$\EF \lesssim \Er$, we see that for $\EF \gtrsim \Er$ one should expect a decrease in $\Gamma$ as we increase the strain $\varepsilon$.
	Assuming that the resonance peak is nearly-symmetrical, we conclude that
	the same estimate as in (\ref{eqn:kappa-m1}), but taken with
	a negative sign, applies. The result is shown in figure~\ref{fig:4}~(a) by the lower dashed enveloping curve.
	
	However, this result should be treated with care. Apart from
	simplicity of the impurity model, it was assumed that the value of the Fermi energy remains constant in the strained sample. As follows from equations (\ref{eqn:dos-0}), (\ref{eqn:w-s}),
	the density of states $\rho_\varepsilon(E)$ changes as we stretch the
	graphene sheet. In an
	isolated sample, the number of charge carriers remains constant. Accordingly, the Fermi energy $\EF^\varepsilon$ becomes strain dependent.
	Neglecting the effects of impurities, the concentration of charge carriers per unit cell reads $\Nc = (\EF^\varepsilon/W_\varepsilon)^2$.
	This gives us
	\begin{equation}
	\EF^\varepsilon \approx \left(1 - \frac{1}{2}\beta \varepsilon \right) \EF.
	\label{eqn:ef-s}
	\end{equation}
	For a positive strain $\varepsilon$, we expect the Fermi energy to reduce to approximately
	$98 \%$ of its initial value per $1 \%$ of strain.
	The Drude weight $D \approx \EF^\varepsilon$ changes accordingly.
	
	The results for constant carrier density are shown in 
	figure~\ref{fig:4}~(b). We present a plot in terms of
	$W_0 \sqrt{N_{\text c}}$, which corresponds to the zero-strain value of the Fermi energy.
	The inclusion of the $\EF^\varepsilon$ shift damps the gain $\varkappa$
	significantly.
	
	Analogously to (\ref{eqn:kappa-m1}), we can estimate the maximum gain
	by including both the effect of the narrowing of the bandwidth (\ref{eqn:w-s}) [already included in equation (\ref{eqn:kappa-m1})]
	and the reduction of the Fermi energy (\ref{eqn:ef-s}). We end up with the following analytical expression:
	\begin{equation}
	\varkappa_{\text m}^{(1+2)} \approx  \dfrac{\beta}{\piup} 
	\left[ \ln \left(\frac{W_0}{|\EF|}\right)+\frac{1}{2} \right].
	\label{eqn:kappa-m}
	\end{equation}
	We can see in figure~\ref{fig:4}~(b) that this estimate
	is approached closely in precise calculation, both for upper and lower enveloping curves.
	
	\subsection{Moderate concentrations} \label{subsec:4.3}
	\begin{figure}[!b]
		\centering
			\includegraphics[width=0.54\textwidth]{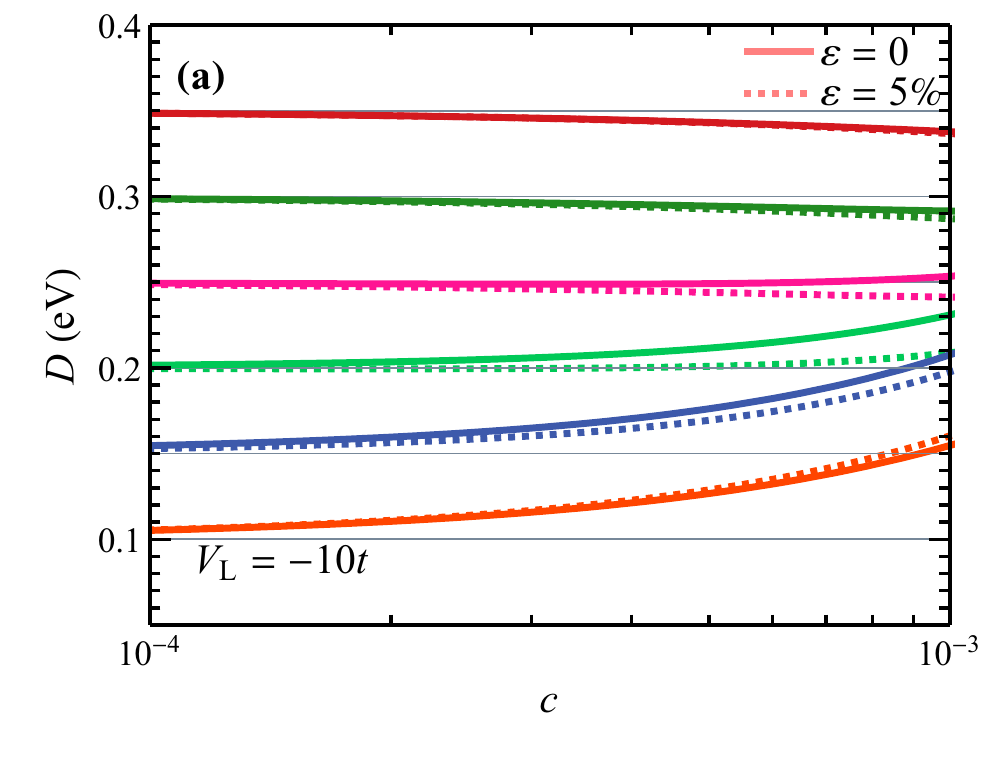}
			\includegraphics[width=0.42\textwidth]{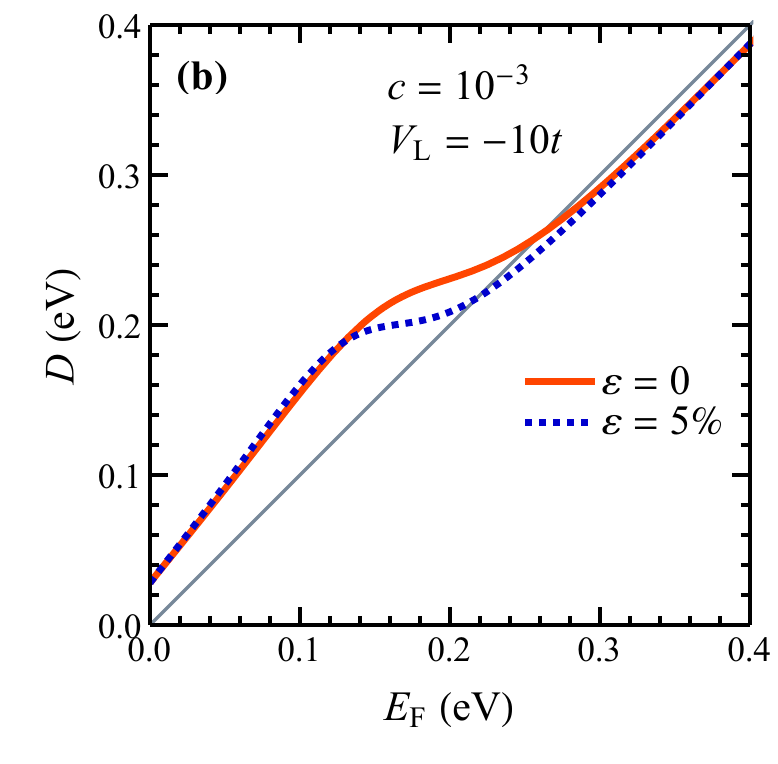}
		\caption{(Colour online) The Drude weight $D$ for $\VL = -10 t$ and two values of strain $\varepsilon = 0$ and $\varepsilon = 5\%$
		(a) as a function of $c$ (logarithmic scale) for various values of $\EF$; (b) as a function of $\EF$ for $c = 10^{-3}$.}
		\label{fig:5}
	\end{figure}
	
	As the impurity concentration increases, we should pay attention to the deviation of the actual Drude weight from its zeroth-order expansion in the impurity concentration $D \approx \EF$.
	Figure~\ref{fig:5}~(a) shows $D(\EF)$ calculated from equation (\ref{eqn:d}) as a function of the concentration $c$ for various values of $\EF$. This deviation increases with $c$, and it is most pronounced for $\EF$ near the resonance energy. The sign of the deviation changes when the Fermi energy crosses the resonance energy. Figure~\ref{fig:5}~(b) shows the dependence of $D$ on the Fermi energy for a fixed concentration $c = 10^{-3}$.
	
	This difference can be understood using the first order expansion (\ref{eqn:d-simple}). As we increase the impurity concentration, the absolute value of $\Re \varSigma(\EF)$ becomes comparable to $\EF$ near the resonance. The real part $\Re \varSigma(\EF)$ changes its sign at the resonance energy (see figure~\ref{fig:sigma}), so $D < \EF$ for $\EF$ smaller than the resonance energy, and vice versa.
	
	We note that $D = D(\EF)$ has a plateau in the vicinity of the resonance energy, as can be seen in figure~\ref{fig:5}~(b). This plateau becomes flatter as we increase the impurity concentration $c$, so one could expect a slow change in $D$ with a large change of $\EF$. This means that one cannot properly determine the Fermi energy~$\EF$ in the resonance region by measuring the Drude weight $D$.
	
	When exceeding certain impurity concentration $c_\chi$, we expect the slope of the $D = D(\EF)$ curve to become negative. We end up with a non-monotonous function, so one cannot determine $\EF$ from $D$ uniquely near the resonance. The curves for different $\EF$ in figure~\ref{fig:5}~(a) cross each other as we approach this concentration. However, for concentrations $c \sim c_\chi$, the applicability criterion for the ATA does not hold. 
	
	We can determine $c_\chi$ by using the first-order expansion in concentration (\ref{eqn:d-simple}). The corresponding estimation can be obtained from the condition of flatness of the function $D = D(E)$ near the solution of the Lifshitz equation $\Er$:
	\begin{equation}
	\left.\frac{\partial}{\partial E} \left[E - \Re \varSigma(E)\right] \right\vert_{E = \Er} = 0 \quad \text{for} \quad c = c_\chi.
	\end{equation}
	This gives us the following estimate:
	\begin{equation}
	c_1 =  \frac{\piup^2}{2} \frac{\Er^2}{W_0^2} \frac{1}{ \ln (W_0/|\Er|)-1}.
	\end{equation}
	For larger values of the perturbation $\VL$, the value of critical concentration is smaller. 
	One can compare it to the spectrum transformation concentration $c_\mathrm{ST}$ \cite{Skrypnyk2007FNT}:
	\begin{equation}
	c_\mathrm{ST} = 2 \frac{\Er^2}{W_0^2} \ln\left(\frac{W}{|\Er|}\right).
	\end{equation}
	
	As we apply the strain, the resonance shifts towards the Dirac point. The dependence $D = D(\EF)$ reflects this shift, as one can see in figure~\ref{fig:5}~(b). Thus, given that $\EF$ remains constant, one should expect the Drude weight $D$ to change. This difference in $D$ for $\varepsilon = 0$ and $\varepsilon = 5\%$ can be seen in figures~\ref{fig:5}~(a), \ref{fig:5}~(b). As we can see in figure~\ref{fig:5}~(b), the most significant difference is observed when the Fermi energy~$\EF$ approaches the resonance energy. However, the Drude weight shift fades as we move away from the resonance. In addition to that, the apparent flatness of the curve near the resonance increases with increasing the strain.
	
	The results of the previous subsection still hold, but one should not assume that $D = \EF$, as it is commonly accepted for the clean graphene. For $\EF$ lower than the resonance energy, where we expect the gain in the Drude width $\Gamma$ (i.e., $\varkappa > 0$), the Drude weight considerably exceeds the corresponding value of the Fermi energy [see figure~\ref{fig:5}~(b)]. Consequently, the values of $D$ for which the maximum of $\varkappa$ is reached will exceed the corresponding values of $\EF$. 

	\section{Conclusion}
	In this paper, the influence of impurities on the Drude peak parameters was studied. Using the Lifshitz model and the average T-matrix approximation, we studied the self-energy function to establish the resonant behaviour of the partial scattering rate for point defects, treated as a function of the Fermi energy. 
	
	Modelling the uniform uniaxial strain by modification of the lattice hopping integrals, we have found that the effect of the strain on the electronic spectrum can be described by introduction of the effective bandwidth. The corresponding bandwidth variation was shown to lead to the shift in the position of the impurity resonance towards the Dirac point with increasing  magnitude of strain. Although this shift is rather moderate in absolute value compared to the resonance energy, it is shown to significantly increase the Drude width. Thus, the Drude peak has a resonance behaviour as a function of Fermi energy. This effect is expressed more pronouncedly for the resonances which are located closer to the Dirac point, i.e., for larger impurity potentials.
	
	In addition, it was shown that the Drude weight is not always a linear function of the Fermi energy. In the extreme case, the Drude weight can have a plateau feature as a function of the Fermi energy, which hinders the determination of the Fermi energy from the Drude weight data. Furthermore, the shift in the resonance energy due to a strain leads to a slight shift in the Drude weight at the constant Fermi energy. The magnitude and the direction of a shift was connected to the impurity resonance energy behaviour.
	
	Although the article \cite{Chhikara2017} gave us an impetus to write the present paper, we did not pose the goal to provide an exhaustive explanation of the experimental results presented in it. This was a reason behind the choice for impurity model. Instead, on a simple example, we made an attempt to show that for an impure graphene, the resonant behaviour of the electronic self-energy function leads to the resonant dependency of the Drude weight on the Fermi energy, and to the subsequent intense change of the Drude weight due to the applied strain.

	\section*{Acknowledgements}
	We are indebted to A.~Kuzmenko for bringing to our attention his experimental results and for illuminating discussions. We are also 
	thankful to V.M.~Loktev for extensive and fruitful discussions.
	S.G.Sh. and Y.V.S. acknowledge a partial support by the National Academy of
	Sciences of Ukraine (project No. 0117U000236) and by its Program of
	Fundamental Research of the Department of Physics and Astronomy (project No.
	0117U000240).

\newpage
	
	\ukrainianpart
	
\title{Спричинене домішками уширення піку Друде у деформованому графені}
\author{В.О. Шубний\refaddr{label1}, С.Г. Шарапов\refaddr{label2}, Ю.В. Скрипник\refaddr{label2}}
\addresses{
	\addr{label1} Фізичний факультет, Київський національний університет імені Тараса Шевченка,\\
	просп. Академіка Глушкова, 6, 03680 Київ, Україна
	\addr{label2} Інститут теоретичної фізики імені М.М. Боголюбова НАН України,\\ вул. Метрологічна, 14-б, 03680 Київ, Україна
}

	\makeukrtitle
	
	\begin{abstract}
		\tolerance=3000%
		Нещодавні експериментальні дослідження далеко-інфрачервоного спектру пропускання в одношаровому графені показали, що ширина піку Друде збільшується більше ніж на 10\% для 1\% прикладеної механічної деформації, при цьому вага піку Друде залишається незмінною. Ми досліджуємо вплив деформації на розсіяння на резонансних домішках. Ми пропонуємо механізм збільшення оберненого часу розсіяння через зсув резонансу. Використовуючи модель Ліфшиця домішок заміщення, ми передбачаємо зміни у ширині і вазі піка Друде як функцій енергії Фермі, концентрації домішок та величини потенціалу домішок.
		\keywords графен, деформації, оптична провідність, точкові дефекти, ширина піку Друде
		
	\end{abstract}

\begin{thebibliography}{99}

\bibitem{Stasyuk}
Stasyuk I.V., Green’s Functions in Quantum Statistics of Solid State: a
Textbook, Ivan Franko National University of Lviv, Lviv, 2013, (in Ukrainian).

\bibitem{Lee2008}
Lee C., Wei X., Kysar J.W., Hone J., Science, 2008, \textbf{321}, No. 5887,
  385--388, \doi{10.1126/science.1157996}.

\bibitem{CastroNeto2009}
{Castro Neto} A.H., Guinea F., Peres N.M.R., Novoselov K.S., Geim A.K., Rev.
  Mod. Phys., 2009, \textbf{81}, No.~1, 109--162,
  \doi{10.1103/RevModPhys.81.109}.

\bibitem{Pellegrino2010prb}
Pellegrino F.M.D., Angilella G.G.N., Pucci R., Phys. Rev. B, 2010, \textbf{81},
  No.~3, 035411,\\ \doi{10.1103/PhysRevB.81.035411}.

\bibitem{Pereira2010epl}
Pereira V.M., Ribeiro R.M., Peres N.M.R., {Castro Neto} A.H., EPL, 2010, \textbf{92}, No.~6, 67001,\\ \doi{10.1209/0295-5075/92/67001}.

\bibitem{Pellegrino2011prb}
Pellegrino F.M.D., Angilella G.G.N., Pucci R., Phys. Rev. B, 2011, \textbf{84},
  No.~19, 195407,\\ \doi{10.1103/PhysRevB.84.195407}.

\bibitem{Pellegrino2011hpr}
Pellegrino F.M., Angilella G.G., Pucci R., High Pressure Res., 2011, \textbf{31},
  No.~1, 98--101,\\ \doi{10.1080/08957959.2010.525705}.

\bibitem{Oliva-Leyva2014jpcm}
Oliva-Leyva M., Naumis G.G., J. Phys.: Condens. Matter, 2014, \textbf{26},
  No.~12, 125302,\\ \doi{10.1088/0953-8984/26/12/125302}.

\bibitem{Ni2014}
Ni G.X., Yang H.Z., Ji W., Baeck S.J., Toh C.T., Ahn J.H., Pereira V.M.,
  \"Ozyilmaz B., Adv. Mater., 2014, \textbf{26}, No.~7, 1081--1086,
  \doi{10.1002/adma.201304156}.

\bibitem{Chhikara2017}
Chhikara M., Gaponenko I., Paruch P., Kuzmenko A.B., 2D Mater., 2017,
  \textbf{4}, No.~2, 025081,\\ \doi{10.1088/2053-1583/aa6c10}.

\bibitem{Ni2010}
Ni Z.H., Ponomarenko L.A., Nair R.R., Yang R., Anissimova S., Grigorieva I.V.,
  Schedin F., Blake~P., Shen~Z.X., Hill~E.H., Novoselov K.S., Geim A.K., Nano
  Lett., 2010, \textbf{10}, No.~10, 3868--3872, \doi{10.1021/nl101399r}.

\bibitem{Andrei2012RPP}
Andrei E.Y., Li G., Du X., Rep. Prog. Phys., 2012, \textbf{75}, No.~5,
  056501, \doi{10.1088/0034-4885/75/5/056501}.

\bibitem{Landau7}
Landau L.D., Pitaevskii L.P., Kosevich A.M., Lifshitz E.M., Theory of
  Elasticity, 3rd~Edn., Butterworth-Heinemann, Oxford, 2012.

\bibitem{Suzuura2002}
Suzuura H., Ando T., Phys. Rev. B, 2002, \textbf{65}, No.~23, 235412,
  \doi{10.1103/PhysRevB.65.235412}.

\bibitem{Naumis2017RPP}
Naumis G.G., Barraza-Lopez S., Oliva-Leyva M., Terrones H., Rep. Prog.
  Phys., 2017, \textbf{80}, No.~9, 096501, \doi{10.1088/1361-6633/aa74ef}.

\bibitem{DeJuan2013a}
{De Juan} F., Ma{\~{n}}es J.L., Vozmediano M.A.H., Phys. Rev. B, 2013,
  \textbf{87}, No.~16, 165131,\\ \doi{10.1103/PhysRevB.87.165131}.

\bibitem{Shah2013}
Shah R., Mohiuddin T.M.G., Singh R.N., Mod. Phys. Lett. B, 2013, \textbf{27},
  No.~03, 1350021,\\ \doi{10.1142/S0217984913500218}.

\bibitem{Lifshitz1964}
Lifshitz I.M., Adv. Phys., 1964, \textbf{13}, No.~52, 483--536,
  \doi{10.1080/00018736400101061}.

\bibitem{Yonezawa1968}
Yonezawa F., Prog. Theor. Phys., 1968, \textbf{40}, No.~4, 734--757,
  \doi{10.1143/PTP.40.734}.

\bibitem{Elliott1974}
Elliott R.J., Krumhansl J.A., Leath P.L., Rev. Mod. Phys., 1974, \textbf{46},
  No.~3, 465--543,\\ \doi{10.1103/RevModPhys.46.465}.

\bibitem{Ehrenreich1976}
Ehrenreich H., Schwartz L.M., Solid State Phys., 1976,
  \textbf{31}, 149--286, \doi{10.1016/S0081-1947(08)60543-3}.

\bibitem{Pershoguba2009}
Pershoguba S.S., Skrypnyk Y.V., Loktev V.M., Phys. Rev. B, 2009, \textbf{80},
  No.~21, 214201,\\ \doi{10.1103/PhysRevB.80.214201}.

\bibitem{LifshitsGredeskulPastur1982}
Lifshits I.M., Gredeskul S.A., Pastur L.A., Introduction to the Theory of
  Disordered Systems, Wiley, New York, 1988.

\bibitem{Kumazaki2006}
Kumazaki H., Hirashima D.S., J. Phys. Soc. Jpn., 2006, \textbf{75}, No.~5,
  053707, \doi{10.1143/JPSJ.75.053707}.

\bibitem{Skrypnyk2007FNT}
Skrypnyk Y.V., Loktev V.M., Fiz. Nizk. Temp., 2007, \textbf{33}, No.~9,
  1002--1007 (in Russian), [Low Temp. Phys., 2007, \textbf{33}, 762, \doi{10.1063/1.2780170}].  

\bibitem{26}
 Skrypnyk Y.V., Loktev V.M., Phys. Rev. B, 2011, \textbf{83}, 085421, \doi{10.1103/PhysRevB.83.085421}.

\end{thebibliography}
\end{document}